\documentclass[journal=mamobx,manuscript=article,layout=twocolumn]{achemso}
\usepackage{multirow}
\usepackage{lineno}
\usepackage[font=footnotesize]{subfig}
\usepackage[format=plain,labelfont={it, bf},textfont=it]{caption}
\usepackage{amsmath,amssymb}
\usepackage{mathtools}
\usepackage{graphicx}
\usepackage[utf8]{inputenc}
\usepackage[english]{babel}
\usepackage{titlesec}
\usepackage{epstopdf}
\usepackage{blindtext}
\usepackage{subfig}
\usepackage{amsmath}
\usepackage{gensymb}
\usepackage{xr}
\makeatletter
\newcommand*{\addFileDependency}[1]{
\typeout{(#1)} 
\@addtofilelist{#1}
\IfFileExists{#1}{}{\typeout{No file #1.}}}
\makeatother
\newcommand*{\myexternaldocument}[1]{%
\externaldocument{#1}%
\addFileDependency{#1.tex}%
\addFileDependency{#1.aux}%
}
\myexternaldocument{suppinfo_black}
\usepackage[T1]{fontenc}
\usepackage[dvipsnames]{xcolor}

\author{Anna Dmochowska}
\affiliation[PIMM]{Laboratoire PIMM, CNRS, Arts et Métiers Institute of Technology, Cnam, HESAM Universite, 75013 Paris, France}
\author{Jorge Peixinho}
\affiliation[PIMM]{Laboratoire PIMM, CNRS, Arts et Métiers Institute of Technology, Cnam, HESAM Universite, 75013 Paris, France}
\email{jorge.peixinho@cnrs.fr}
\author{Cyrille Sollogoub}
\affiliation[PIMM]{Laboratoire PIMM, CNRS, Arts et Métiers Institute of Technology, Cnam, HESAM Universite, 75013 Paris, France}
\author{Guillaume Miquelard-Garnier}
\affiliation[PIMM]{Laboratoire PIMM, CNRS, Arts et Métiers Institute of Technology, Cnam, HESAM Universite, 75013 Paris, France}
\email{guillaume.miquelardgarnier@lecnam.net}

\title{Extensional Viscosity of Immiscible Polymers Multinanolayer Films: Signature of the Interphase\footnote{This document is the unedited Author's version of a Submitted Work that was subsequently accepted for publication in Macromolecules, copyright © 2023 American Chemical Society after peer review. To access the final edited and published work see https://doi.org/10.1021/acs.macromol.3c00288}}


\begin{document}

\begin{tocentry}
\includegraphics[width=1.0\columnwidth,keepaspectratio]{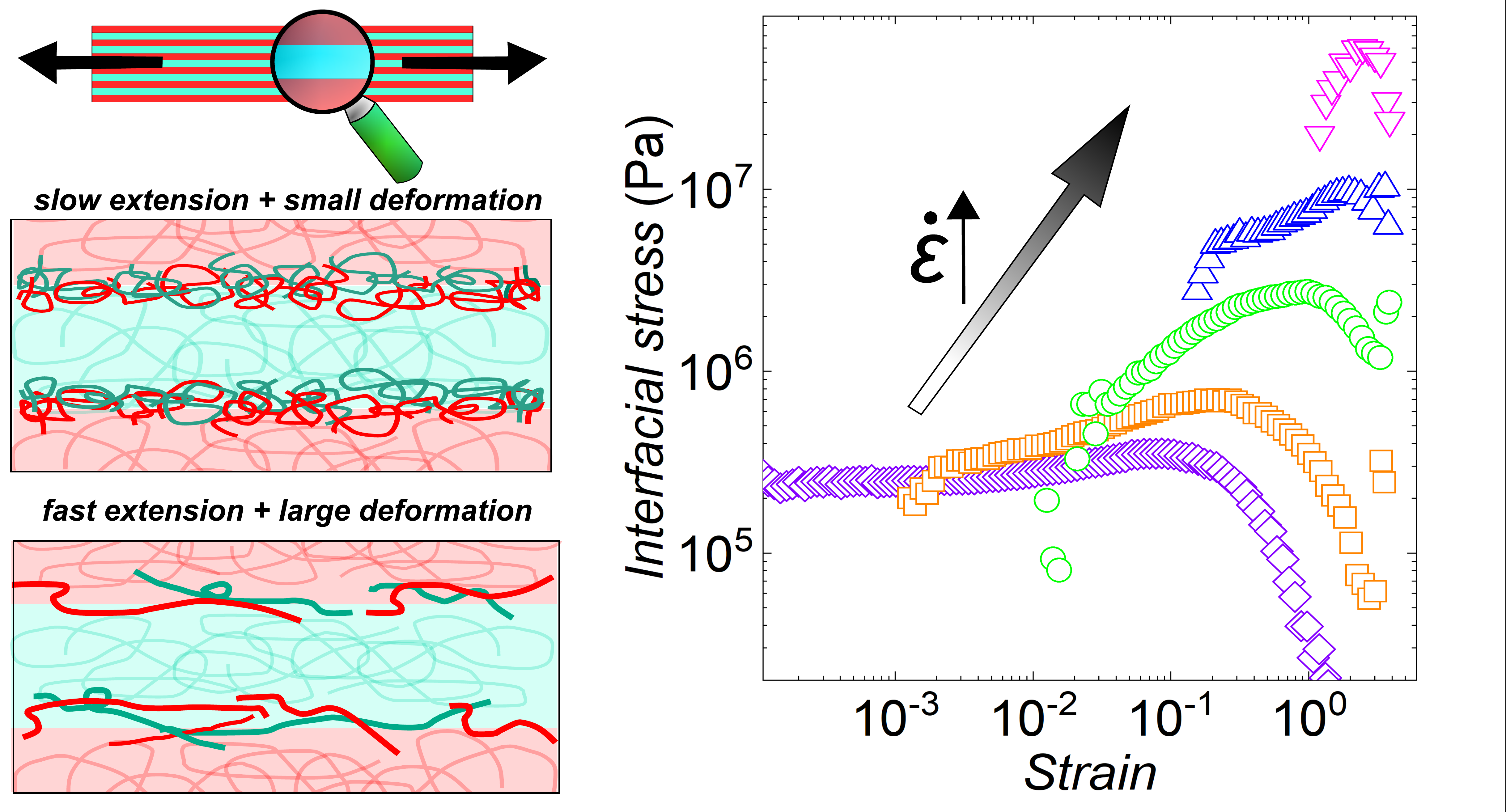}
\end{tocentry}

\section*{Abstract}
The measurement of interfacial mechanical or rheological properties in polymer blends is a challenging task, as well as providing a quantitative link between these properties and the interfacial nanostructure.
Here, we perform a systematic study of the extensional rheology of multilayer films of an immsicible polymer pair, polystyrene and poly(methyl methacrylate). We take advantage of multinanolayer coextrusion to increase the number of interfaces up to thousands, consequently magnifying the interfacial response of the films.
The transient elongational response is compared to an addivity rule model based on the summation of the contribution of each polymer as well as the interfacial one. 
At low strain rates, the model reproduces the transient extensional viscosity up to strain-thinning, while at larger ones, the extra stress exceeds the prediction based on constant interfacial tension. This extra-contribution is attributed to an interphase modulus on the order of 1-10 MPa, which increases with strain rate following a power-law with an exponent 1/3. Extensional rheology of multinanolayer films is then an efficient combination to go beyond interfacial tension and measure quantitatively the interfacial rheology of immiscible polymer blends. 

\section*{Introduction}

Polymer blends represent today more than a third of the world's plastic production, despite the fact that most polymers are immiscible\cite{Utracki1998}, which  results in phase separation and unentangled interfaces. Compatibilization has been mastered for a long time industrially to achieve suitable properties for such immiscible blends. It consists of modifying the interface with several possible strategies, such as the addition or the in-situ formation of a copolymer\cite{Formela2018} that will segregate at the interface and allow entanglements, the use of nanoparticles\cite{Kwon2011,Salzanodeluna2016,Qiao2023}, cosolvent\cite{ZhangG2016}, or through ionic bonds\cite{Fredrickson2022} and electrostatic correlations\cite{Benmouna1991}.

The effect of the compatibilizer, such as the molecular weight or the amount of copolymer, has been quantified long ago on interfacial properties in the solid state such as fracture toughness at the interface\cite{Creton1991,Creton1992}. However, it was only recently that the rheological properties of compatibilized interfaces have been fully characterized\cite{Beuguel2019,Beuguel2020}. To do so, we took advantage of multinanolayer coextrusion that applies successive slicing and recombining a stratified polymer melt flow giving rise to a material made of thousands of alternating nanometric layers\cite{Li2020,Baer2017}. By doing so, the effect of interfaces was drastically enhanced in the response of the materials under oscillatory shear.
In non-compatibilized blends, a quantitative link has also been established between the fracture toughness and, this time, the interfacial nanometric thickness, which depends on the Flory interaction parameter between the two polymers. Similarly, in the melt, a relation between viscoelastic moduli and interfacial tension has been proposed \cite{Palierne1990,Graebling1993} and used to describe the melt properties of polymer blends. Still, measuring interfacial tension of immiscible polymers is a tedious task\cite{Xing2000}, due to the high viscosities and temperatures involved, and does not always provide sufficient information: even in simple mixtures, flows may be impacted by surface tension gradients (the well-known Marangoni effect). Hence, a full characterization of the interfacial rheology, as well as its evolution in relation with the morphology of the interface and its impact on the processing of nanostructured blends,  is lacking in the literature. 

Extensional rheology is a challenging but rapidly developing technique to measure elongational flows, which allowed in particular to reexamine fundamental theories in polymer physics, such as the tube model\cite{Doi1988,Huang2022}.
In this study, we aim at taking advantage of this technique applied to multinanolayer films of a well-known polymer couple, polystyrene (PS) and poly(methyl methacrylate) (PMMA), in order to study the flow properties of their interface. Extensional rheology of PS \cite{Andrade2014,Huang2015} and PMMA\cite{Stamboulides2006,Morelly2019} has already been investigated separately. The effect of molecular weight, a rise of transient extensional viscosity above the predictions from linear viscoelasticity theory and strain hardening have been well-documented for both polymers.
However, there has been only a few reports of extensional rheometry of multilayer films. Notably, Lamnawar et al. have studied the dynamics of interdiffusion across the interfaces during processing of a miscible polymer pair\cite{Zhang2012,Zhang2016,Lu2018,Zhao2021}, as well as in-situ compatibilization reaction of an immiscible one \cite{Lu2020}. A more model approach by Macosko et al. aimed at tackling the interfacial contribution in the rheological response of multilayer films of various polymer pairs, coupling experiments and a model based on additivity rule (\textit{i.e.} a viscosity that is the arithmetic average of the viscosity of each phase)\cite{Levitt1997,Jordan2019}. 
Building on this approach, we propose here a systematic study of the transient extensional stress of PS/PMMA multinanolayer films, where the number of interfaces is varied from 2 to 4096 and the strain rates from 0.001 to 10 s$^{-1}$, hence from quasi-static to non-linear flow. Such high number of interfaces increases drastically the interfacial contribution in the rheological response of the films and allows the measurement of its extensional viscosity and its comparison with theoretical predictions. An increase of the measured stress at high strain, dependent of strain rate, is evidenced, and is reminiscent of the surface elasticity in soap films.

\section*{Materials and Methods}
\subsection*{Materials}
Polystyrene PS 1340 from Total and poly(methyl methacrylate) PMMA VM100 from Arkema were selected to produce multilayer films, based on an earlier work \cite{Bironeau2017}. The molecular weights, glass transition temperatures and densities have been determined previously \cite{Bironeau2017,Zhu2016}. The polymers were chosen so that at the coextrusion temperature the viscosity ratio close to 1 is achieved in the range of shear rates applied during coextrusion (Figure S1). The viscoelastic properties of the neat polymers have been obtained by small amplitude oscillatory shear (SAOS) measurements at several temperatures ranging from 130 \textdegree C to 225 \textdegree C with a DHR 20 (TA Instruments) rheometer with a plate-plate geometry (25 mm diameter and 1 mm gap) under air flow and not nitrogen to simulate the coextrusion conditions. Frequency sweep tests were conducted in the range from 0.045 to 628 rad/s with an applied strain of 1 \%, in the linear viscoelasticity regime, and confirmed a comparable thermorheological behavior of the two polymers (Figure S3c) \cite{Plazek1965,Huang2015}. The zero-shear viscosity, $\eta_0$, was determined by a classical Carreau-Yasuda model \cite{Bird1968}, similarly to our previous work \cite{Dmochowska2022}. The main properties of the neat polymers are listed in Table S1. 

\subsection*{Films fabrication}
PS/PMMA films were produced by using a lab-made customized multilayer coextrusion line (Figure \ref{fig:fig1})\cite{Montana2018}.
It consists of three 20 mm single-screw Rheoscam extruders (Scamex), two melt-gear pumps, a three-layer feed block, layer-multiplying elements (LME), a flat die, and a chill roll. The temperature of the PS and PMMA extruders, feed block, and LME assembly was set to 225 \textdegree C. The composition can be controlled by tuning the screw speed and controlling the gear pumps. The polymer flows enter a three-layer coextrusion feed block and next pass through a series of layer-multiplying elements, where the flow for each LME is first split vertically, then spread horizontally and recombined. The total thickness of the flow remains the same throughout the process. An assembly of \emph N LMEs results in a film composed of $n=2^{N+1}+1$ layers. In this study, films with 3, 17, 129, 2049, and 4097 layers were obtained with 0, 3, 6, 10, and 11 LMEs, respectively. After exiting the last LME, the polymer melt enters a flat die with a 2 mm die gap and 150 mm width which temperature is set to 200 \textdegree C. Finally, the film is collected using a chill roll heated up to 90 \textdegree C with the lowest possible drawing speed to prevent post-extrusion chain relaxation. Additionally, to reduce the thickness of the final film without any post-stretching step, a sacrificial layer of low-density polyethylene, LDPE 1022 FN, is added at the exit die.

In this work, the two studied weight compositions of PS/PMMA multilayer films are close to 60/40 and 30/70 and have thicknesses, $H_{\text{\tiny{M}}}$, lower than 1 mm to fulfill requirements of the extensional rheology measurements. The exact compositions and thicknesses of the extruded films are given in Table S2.

\subsection*{Films morphology}
The morphology and individual layer thicknesses of the films were characterized with an optical microscope Axio Imager 2 (ZEISS) or an atomic force microscope (AFM) Nanoscope V (Veeco), depending on the expected layer thickness. In both cases, the samples were cut from the center of the extruded film parallel to extrusion flow and cross-sectioned perpendicular to their surface by using an ultra-microtome with a diamond knife (Diatome). Thickness of at least 10 \% layers was measured for each film by following the procedure developed in Bironeau et al. \cite{Bironeau2016}

\subsection*{Extensional rheology}
The viscoelastic properties in uniaxial stretching were determined by a rheometer MCR 502 (Anton Paar) equipped with Sentmanat Extensional Rheometer platform SER-2 (Xpansion Instruments) \cite{Sentmanat2004,Sentmanat2005}, which consists of paired drums coupled to the motor. The extensional viscosity is proportional to the stretching force related to the torque and the assumption of exponential decay of cross-sectional area of the sample. The films were cut into rectangular samples with length (\emph{L}) about 20 mm and width (\emph{W}) around 11 mm and tested at 155 \textdegree C under air flow. This temperature was optimized so that no thermal degradation will affect the  viscosity of the samples over the experimental timescale (see Figure S2), but also that the shrinking is negligible while the torque is measurable \cite{Costanzo2019}. Five Hencky strain rates ($\dot{\varepsilon}$) ranging from 0.001 to 10 s$^{-1}$, directly proportional to the shaft rotation rate, were chosen for the measurements and kept constant over test time. The films were stretched along the extrusion direction. All measurements were repeated at least three times and averaged. The tests were performed until the sample breakage or, depending on which happens first, until a strain value of 3.8 due to the limitation of one drum revolution in SER system \cite{Maia2014,Hirschberg2023}.

\section*{Results and discussion}

\subsection*{Films morphology}
\begin{figure*}[hbt!]
  \centering
  \includegraphics[width=17.6cm]{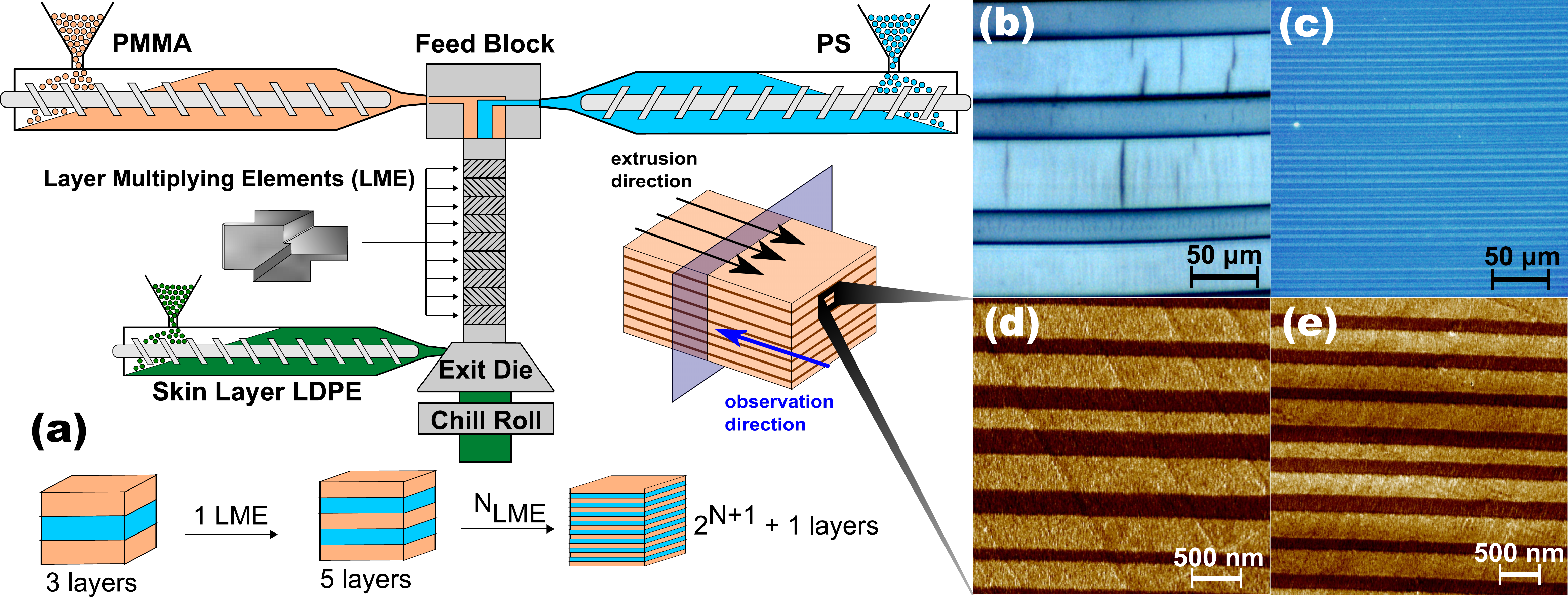}
  \caption{Multinanolayer coextrusion scheme (a) with typical images of fabricated films: (b) 17 layers (the black vertical strokes are compression lines present due to cutting); (c) 129 layers; (d) 2049 layers; (e) 4097 layers. On images from optical microscope (b, c), the lighter blue corresponds to PS, darker blue to PMMA. On AFM images (d, e), the colour gold represents PS layers while brown represents PMMA layers.} 
  \label{fig:fig1}
\end{figure*}

The morphology of fabricated PS/PMMA multilayer films is presented in Figure \ref{fig:fig1}. 
The averaged thicknesses of individual layers of all films are in good agreement with the calculated values obtained from eq S1, as shown in Table S2 in supplementary material.

In all cases, the morphology analysis revealed some heterogeneity in the layer thicknesses of our samples, which is not unusual for such samples\cite{Bironeau2016} and can also be amplified by the fact that we used the lowest possible draw ratio. As expected, no broken layers are observed for films with 3, 17, or 129 layers. For all but one film with 2049 and 4097 layers, the amount of broken layers is less than 5 \%. In one film with 4096 layers and the respective average thicknesses of PS and PMMA equal to 124 nm and 108 nm, the amount of broken layers is close to 9 \%. Still, due to the low amount of broken layers, they will be neglected in the following analyses and we will use the average layer thickness. 

\subsection*{Rheological investigation}
\subsubsection*{Neat polymer melts}

In order to proceed from oscillation to extension, the multimode Maxwell model, eqs \ref{eq1} and \ref{eq2}, was used to determine the linear viscoelastic (LVE) envelope from small-amplitude oscillatory shear (SAOS) data\cite{Bourg2021}. The time temperature superposition of storage, \emph{G}', and loss, \emph{G}'', moduli for PS and PMMA was done at the reference temperature 155 \textdegree C, same as temperature of extensional rheology experiments (see Figure S3a, b). The calculations were done according to the following equations:

\begin{equation}
G'\left(\omega\right) =  \sum_{i=1}^{N}g_i \frac{\left(\omega\tau\right)^2}{1+\left(\omega\tau\right)^2}
\label{eq1}
\end{equation}

\begin{equation}
G''\left(\omega\right) =  \sum_{i=1}^{N}g_i \frac{\omega\tau}{1+\left(\omega\tau\right)^2}
\label{eq2}
\end{equation}
where $g_i$ is the relaxation modulus (in Pa) and $\tau_i$ is the time constant (s). The specific values can be found in Table S3 in supplementary material.

\begin{figure}[hbt!]
  \centering
  \includegraphics[width=8.4cm]{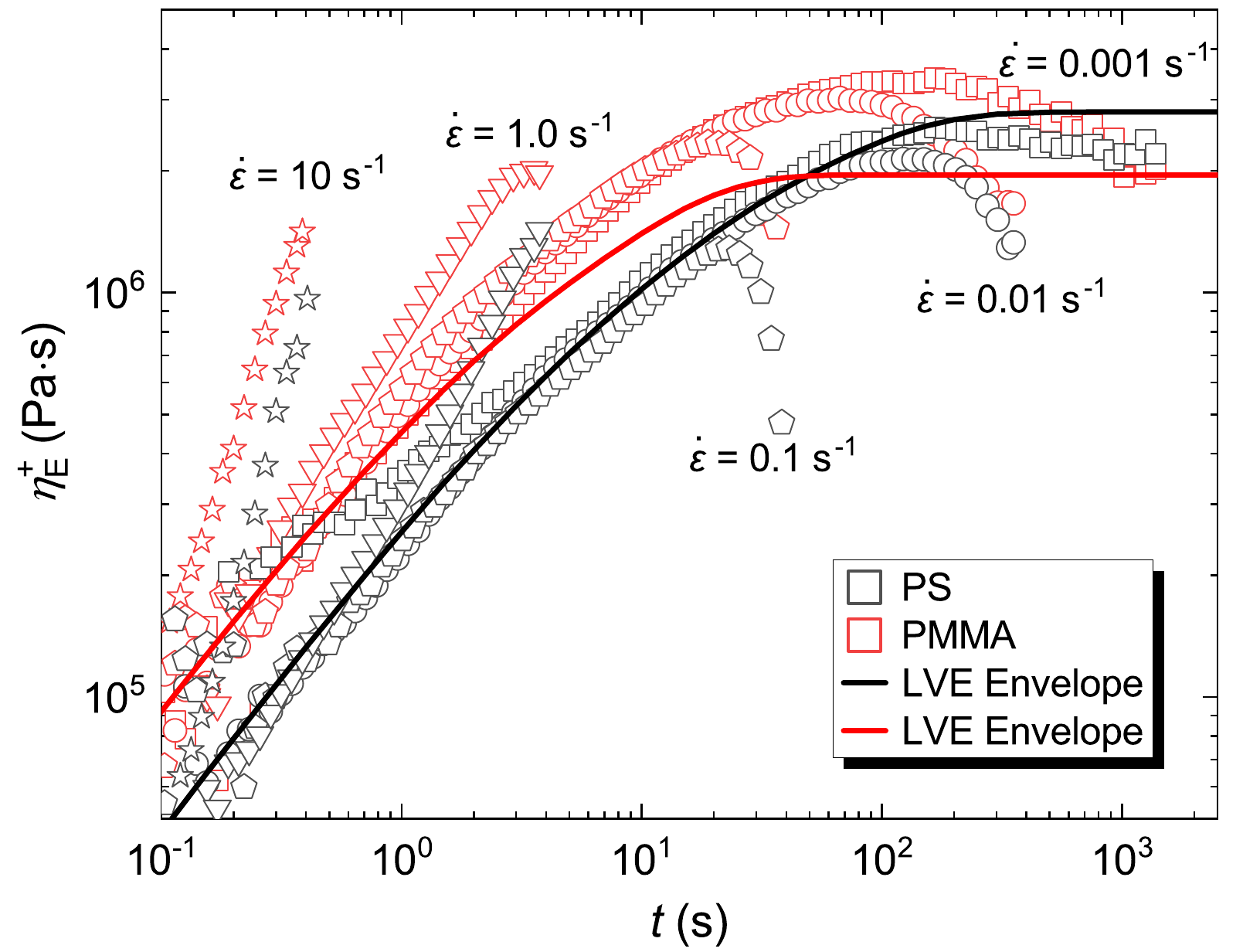}
  \caption{Extensional viscosity of PS and PMMA melts as a function of time, at 155 \textdegree C, at five chosen Hencky strain rates: 0.001, 0.01, 0.1, 1 and 10 s$^{-1}$. The solid lines represent the LVE envelope determined by Maxwell model.} 
  \label{fig:fig2}
\end{figure}

The extensional viscosity in the LVE regime, $\eta_\text{E}$, calculated from SAOS measurements was plotted as a function of time, \emph{t}, according to Trouton's ratio \cite{Barnes1989,Dealy2018} by using the previously found values of relaxation modulus and time constant:

\begin{equation}
\eta_\text{E} \left(t\right) = 3 \sum_{i=1}^{N}g_i\tau_i \left( 1-e^\frac{-t}{\tau_i}\right)
\label{eq3}
\end{equation}

Figure \ref{fig:fig2} presents measured extensional viscosity, $\eta_\text{E}^{+}$,  as a function of time for PS and PMMA at different constant Hencky strain rates, $\dot{\varepsilon}$. The studied polymers exhibit similar viscoelastic properties in SAOS, therefore it is expected that it will be a similar case in extension. As seen on Figure \ref{fig:fig2}, PS at low strain rates (0.001 to 0.1 s$^{-1}$) follows the linear viscoelastic (LVE) envelope at the beginning of the measurement, \textit{e.g.} at lower strain. However, towards higher measurement times, closer to the sample breakage, extensional viscosity decreases in comparison with the LVE values. That could indicate a strain-softening behavior related to chain stiffness \cite{Zhang2016}. Inversely, at high strain rates (1 and 10 s$^{-1}$), it is clearly seen that viscosity overshoots the LVE envelope values from the start of the measurement. This can be explained by the fact that elastic forces overcome the viscous ones, as revealed by a Weissenberg number $W_\text{i}=\dot{\varepsilon}\lambda$ higher than 1\cite{Thompson2021}. Here, $\lambda$ is the terminal relaxation time of PS at 155 \textdegree C (taken from the crossing of \emph{G'} and \emph{G''} in SAOS, see Figure S3) $\approx$ 7 seconds. The measured viscosity increases more rapidly than the LVE and continues to a plateau value outside of the LVE regime\cite{Matsumiya2018,Huang2019}, which was not reached before breakage in the present case.  \\
Similarly to PS, PMMA exhibits a viscosity overshoot at high strain rates. The relaxation time being about 2 s, the strain rate at which $W_\text{i}$ is higher than 1 is 0.5 s$^{-1}$. At lower strain rates, even though PMMA is closer to a regime of a weak linear flow, the LVE envelope is only followed at the beginning of the measurements and a deviation towards higher viscosity values is noted\cite{Rasmussen2019,Stamboulides2006}. In contrast, towards the sample breakage, strain-softening due to chain stiffness \cite{Iisaka1978,Javadi2018} is observed, similarly to PS samples\cite{Morelly2019,Stamboulides2006}.
 
\subsubsection*{Multilayer films}

\begin{figure*}[hbt!]
  \centering
  \includegraphics[width=17.6cm]{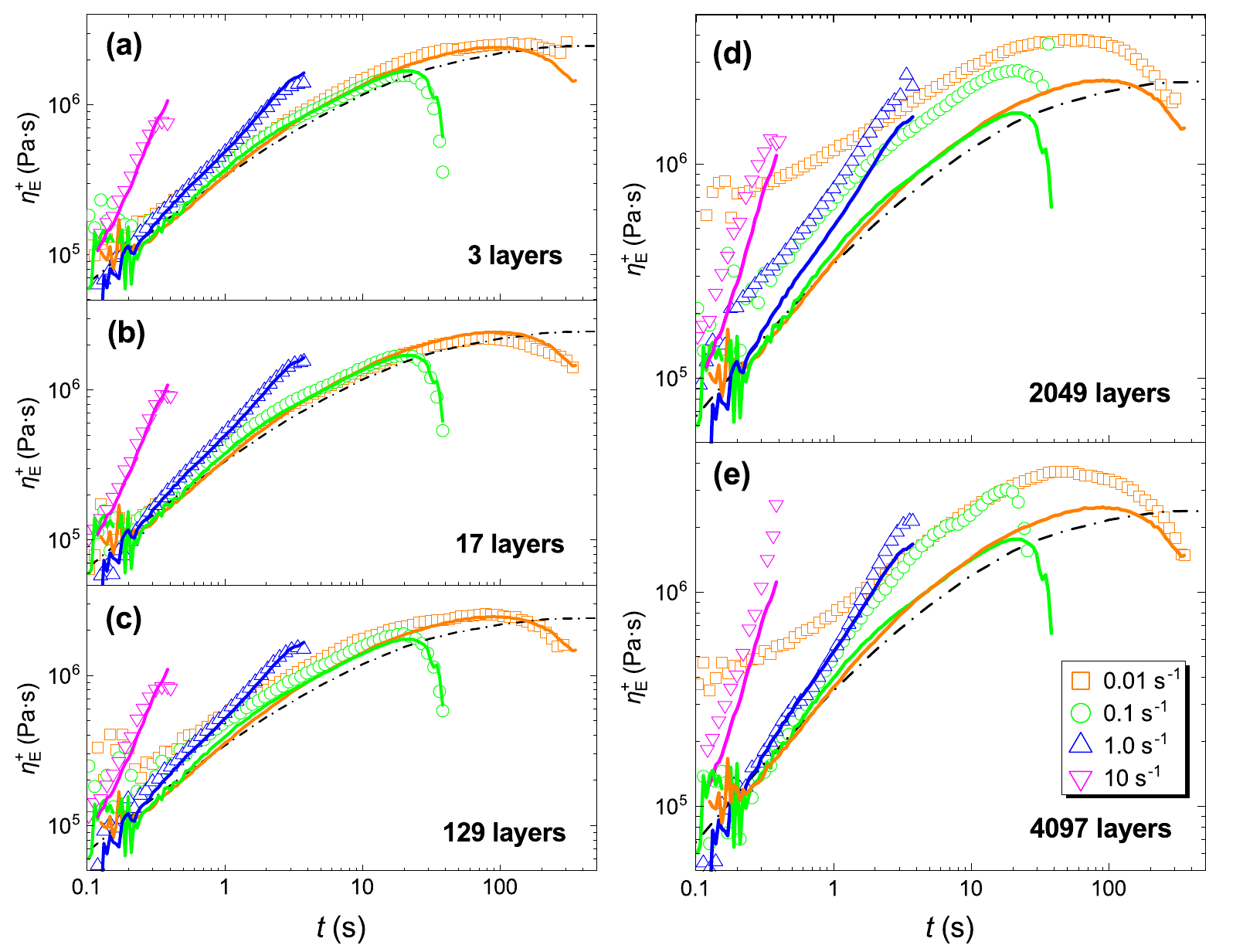}
  \caption{Extensional viscosity for four Hencky strain rates of films with various number of layers: (a) 3, (b) 17, (c) 129, (d) 2049, (e) 4097 layers. The solid lines represent the additivity rule calculated from eq \ref{eq4} for each strain rate and the dashed lines the additivity rule of LVE data. The composition of the presented films is close to 60/40 PS/PMMA in all cases (see Table S2 for details).}
  \label{fig:fig3}
\end{figure*}

Extensional viscosity of the 60/40 PS/PMMA multilayer films with various numbers of layers, $\eta_{\text{\tiny{E,M}}}^+$, is presented as a function of time in Figure \ref{fig:fig3}. In order to understand the behavior of the multilayer films, a theoretical value of their viscosity was calculated at each Hencky strain rate. 
\\ In a first simple approach\cite{Lu2017,Lu2020}, and following the theoretical framework initially developed by Macosko and co-workers \cite{Levitt1997}, we assume that the stress within the multilayer film, $\sigma_{\text{\tiny{E,M}}}^+$, which is proportional to the extensional viscosity and the overall extension rate, follows a simple additivity rule, hence is a sum of stresses gathered within each component, PS layers and PMMA layers, following eq \ref{eq4}:
\begin{equation}
\begin{split}
&\sigma_{\text{\tiny{E,M}}}^+\left(t\right)
=
\eta_{\text{\tiny{E}}}^+\left(t\right)\dot{\varepsilon}
=\\
&\varphi_{\text{\tiny{PS}}}\sigma_{\text{\tiny{E,PS}}}^+\left(t\right)
+ \varphi_{\text{\tiny{PMMA}}}\sigma_{\text{\tiny{E,PMMA}}}^+\left(t\right) 
\label{eq4}
\end{split}
\end{equation}
where $\varphi_{\text{\tiny{PS}}}$ and $\varphi_{\text{\tiny{PMMA}}}$ are the volume fractions of PS and PMMA, respectively. The melt volume fractions of both PS and PMMA were corrected with a melt volume ratio including the variation in density values of the materials at room temperature and experiment temperature (see eq S5).
 
The prediction of LVE envelope for multilayer films was also calculated with a similar approach, using the respective LVE envelopes of PS and PMMA obtained through the Maxwell model.

As presented in Figure \ref{fig:fig3} for 60/40 PS/PMMA composition, this basic additivity rule describes well the experimental data for all strain rates in the case of samples with 3, 17 and 129 layers. All three samples display a similar behavior to PS and PMMA, as expected. At low strain rates, a strain-softening behavior is observed towards the end of the measurement, closer to breakage. At high strain rates, an overshoot from the predicted LVE envelope occurs, similarly to the neat polymer melts, hence well-captured by the additivity rule.
On the other hand, samples with 2049 and 4097 layers display a much different behavior. Starting with the lowest tested strain rate, 0.01 s$^{-1}$, a large increase in the values of $\eta_{\text{\tiny{E,M}}}^+$ compared to the prediction from eq \ref{eq4} is observed from the very beginning of the measurement \cite{Nielsen2009,Andrade2014}. Though less pronounced at higher strain rates, the same observations can be made at all strain rates. Note that similar behavior is observed for the 30/70 PS/PMMA composition (see Figure S4).
In the films with 2049 and 4097 layers, the number of interfaces is, as stated previously, extremely high. Though PS and PMMA display a poor compatibility, their chains will still slightly interpenetrate at the interface, creating, what we will call in the following, an interphase of typical thickness \cite{Helfand1972}:

\begin{equation}
a_{\text{int}} \approx \frac{2b}{\sqrt{6 \chi }} 
\label{eq5}
\end{equation}
where $\chi$ is the Flory-Huggins interaction parameter, and $b$ is the effective length per monomer unit \cite{Helfand1972} for which PS and PMMA have very similar values \cite{Miquelard2016},
$b_{\text{PS}}=6.8$ \text{\AA} \cite{Ballard1973} and  
$b_{\text{PMMA}}=7.4$ \text{\AA} \cite{Kirste1967}. To obtain $\chi$ at 155 \textdegree C, we use the well-known relation proposed by Russell et al. \cite{Russell1990}, leading to $\chi \approx 0.37$, which gives an interphase thickness close to 3 nm. This low value, smaller than the entanglement length, is, as discussed in the introduction, notably responsible for the weak interfacial adhesion between these two polymers in the glassy state\cite{Schnell1998}. 

Still, for films where the layer thicknesses are on the order of 100 nm (see Table S2), the interphase volume fraction becomes non negligible, especially since it will increase during the rheological test which slims down the film.
To take into account this possible effect of an interphase in the extensional viscosity of multilayer films, Macosko and coworkers \cite{Levitt1997,Jordan2019} proposed a refined version of the additivity rule incorporating the interfacial contribution:

\begin{equation}
\begin{split}
\sigma_{\text{\tiny{E,M}}}^+\left(t\right) & = \varphi_{\text{\tiny{PS}}} \sigma_{\text{\tiny{E,PS}}}^+\left(t\right) \\
 & + \varphi_{\text{\tiny{PMMA}}}\sigma_{\text{\tiny{E,PMMA}}}^+\left(t\right)
+ \varphi_{\text{\tiny{int}}}\sigma_{\text{\tiny{E,int}}}^+ \!\left(t\right)   
\label{eq6}
\end{split}
\end{equation}
where $\varphi_{\text{\tiny{int}}}$ and $\sigma_{\text{\tiny{E,int}}}^+ \!\left(t\right)   $ are the volume fraction and the stress of the interphase, respectively (note that taking into account the interphase, the volume fraction of PS and PMMA layers in eqs \ref{eq4} and \ref{eq6} are then slightly different from each other for a given film).

$\varphi_{\text{\tiny{int}}}$ is here simply defined as the total thickness of the interphase (the interphase thickness multiplied by the number of interfaces) divided by the total thickness of the film:

\begin{equation}
\varphi_{\text{\tiny{int}}}=\frac {H_{\text{int}}}{H_{\text{M}}} = \frac{a_{\text{int}} \left( n-1 \right)}{H_{\text{M}}}.
\label{eq7}
\end{equation}
We have to consider that during the measurement, the dimensions of the sample vary with time and strain (see supplementary material, eqs S2-S4). 
Especially, $H_{\text{M}}$ is predicted to decrease exponentially, which we verified experimentally as described in Figures S5 and S6. Therefore, if we make the hypothesis that the interphase typical thickness does not evolve significantly during the test (which will be discussed further later on), then the fraction of interphase will be gradually increasing over the experiment time.
Assuming the chains are not oriented at the interfaces (i.e. for low $W_\text{i}$), the interfacial stress, $\sigma_{\text{\tiny{E,int}}}^+ \!\left(t\right)$, at equilibrium can be related to interfacial tension and interphase thickness through the relation \cite{Jordan2019}: 
\begin{equation}
\sigma_{\text{\tiny{E,int}}}^+ \!\left(t\right) = \frac{\Gamma}{a_{\text{int}}}
\label{eq8}
\end{equation}

Substituting the stress with viscosity, the following equation describing the extensional viscosity of the multilayer film can be obtained \cite{Jordan2019}:
\begin{equation} 
\begin{split}
& \eta_{\text{\tiny{E,M}}}^+\! \left(t\right) = \\
& \varphi_{\text{\tiny{PS}}}\eta_{\text{\tiny{E,PS}}}^+\! \left(t\right)  + \varphi_{\text{\tiny{PMMA}}}\eta_{\text{\tiny{E,PMMA}}}^+\! \left(t\right) 
+ \frac{\Gamma \left(n-1\right)}
{\dot{\varepsilon} H_0 \exp{\left(\frac{-\dot{\varepsilon}\text{t}}{2}\right)}}.
\end{split}
\label{eq9}
\end{equation}

Interfacial tension can be obtained, as can the interfacial thickness, from the theoretical work of Helfand\cite{Helfand1971,Helfand1972}:

\begin{equation}
\Gamma = \frac{kT}{b^2} \left( \frac{\chi}{6} \right)^{1/2}
  \label{eq10}
\end{equation}
with \emph{k} the Boltzmann constant.
The obtained value at the 155 \textdegree C is 0.92 mN/m, similar to the one that can be extrapolated from Wu's experimental work (1.45 mN/m) \cite{Wu1970,Miquelard2016}.

\begin{figure*}[ht!]
  \centering
  \includegraphics[width=8.6cm]{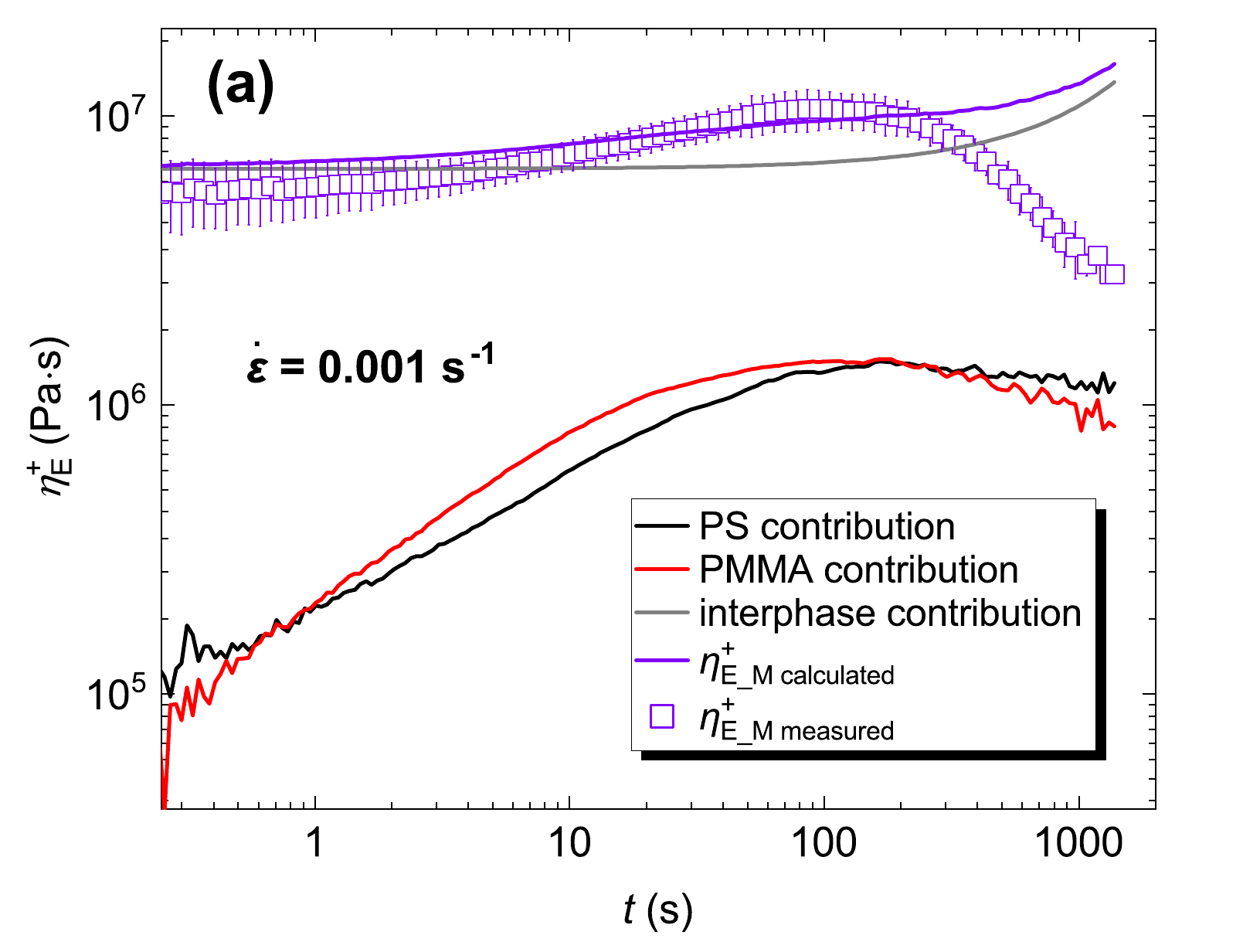}
  \includegraphics[width=8.6cm]{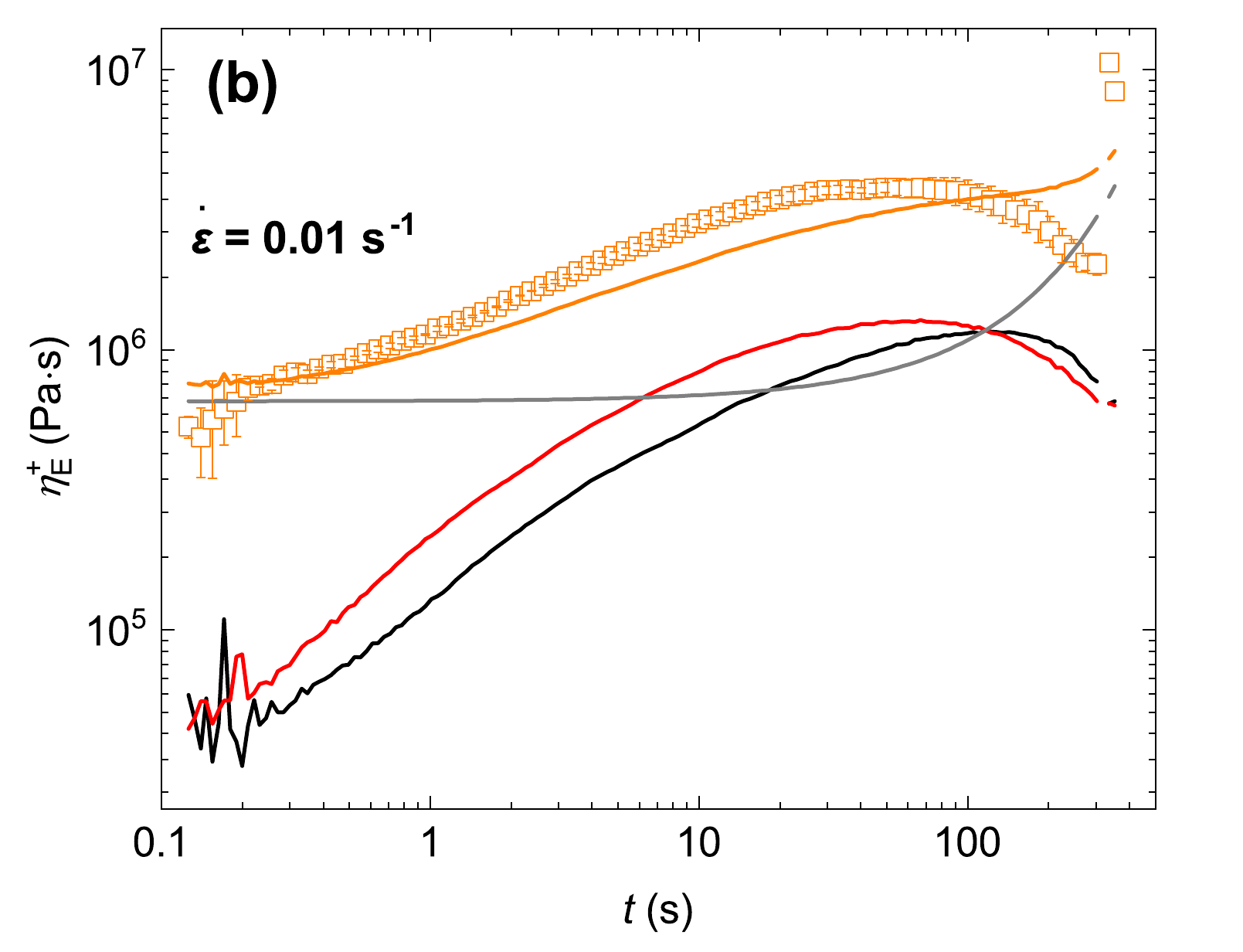}
  \includegraphics[width=8.6cm]{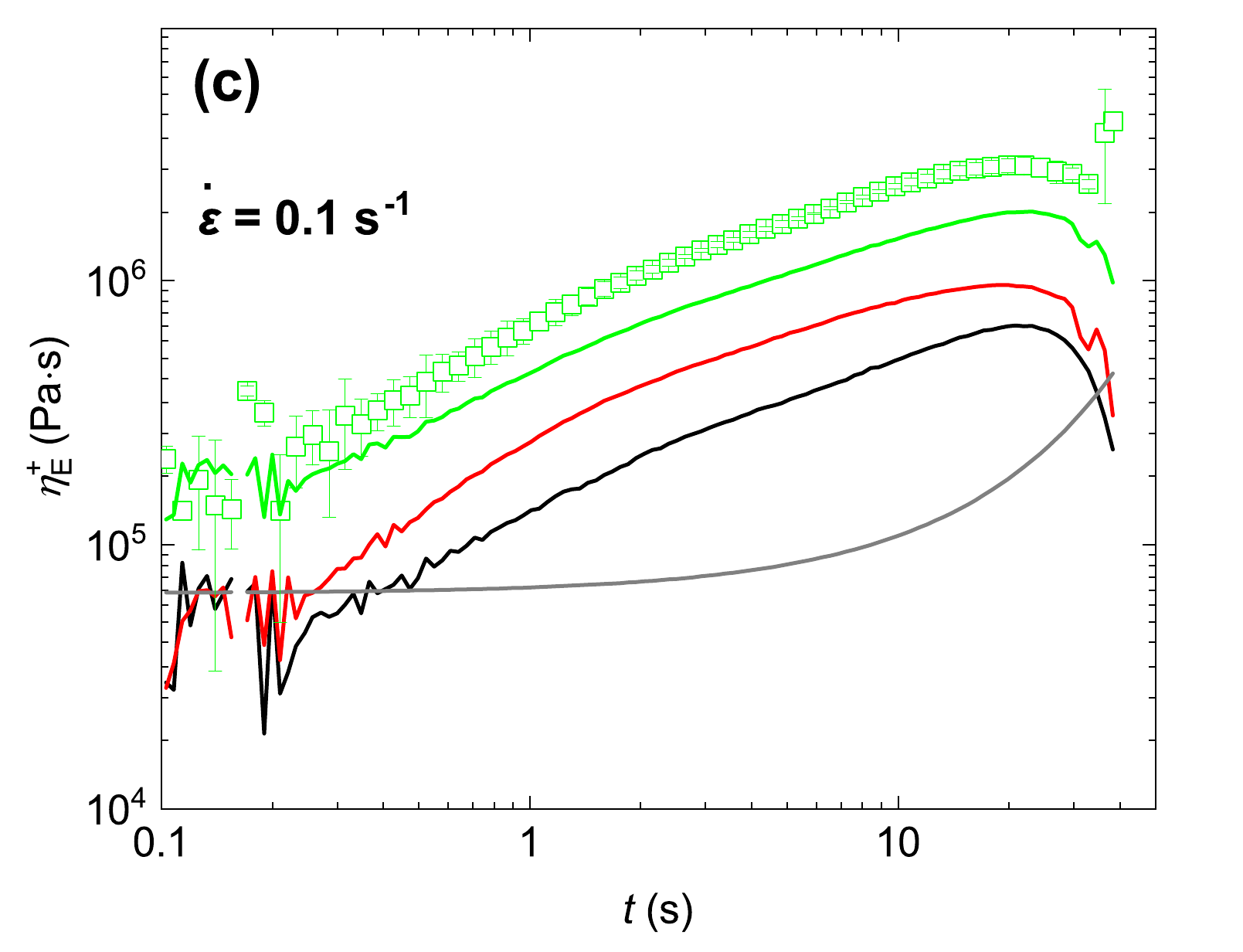}
  \includegraphics[width=8.6cm]{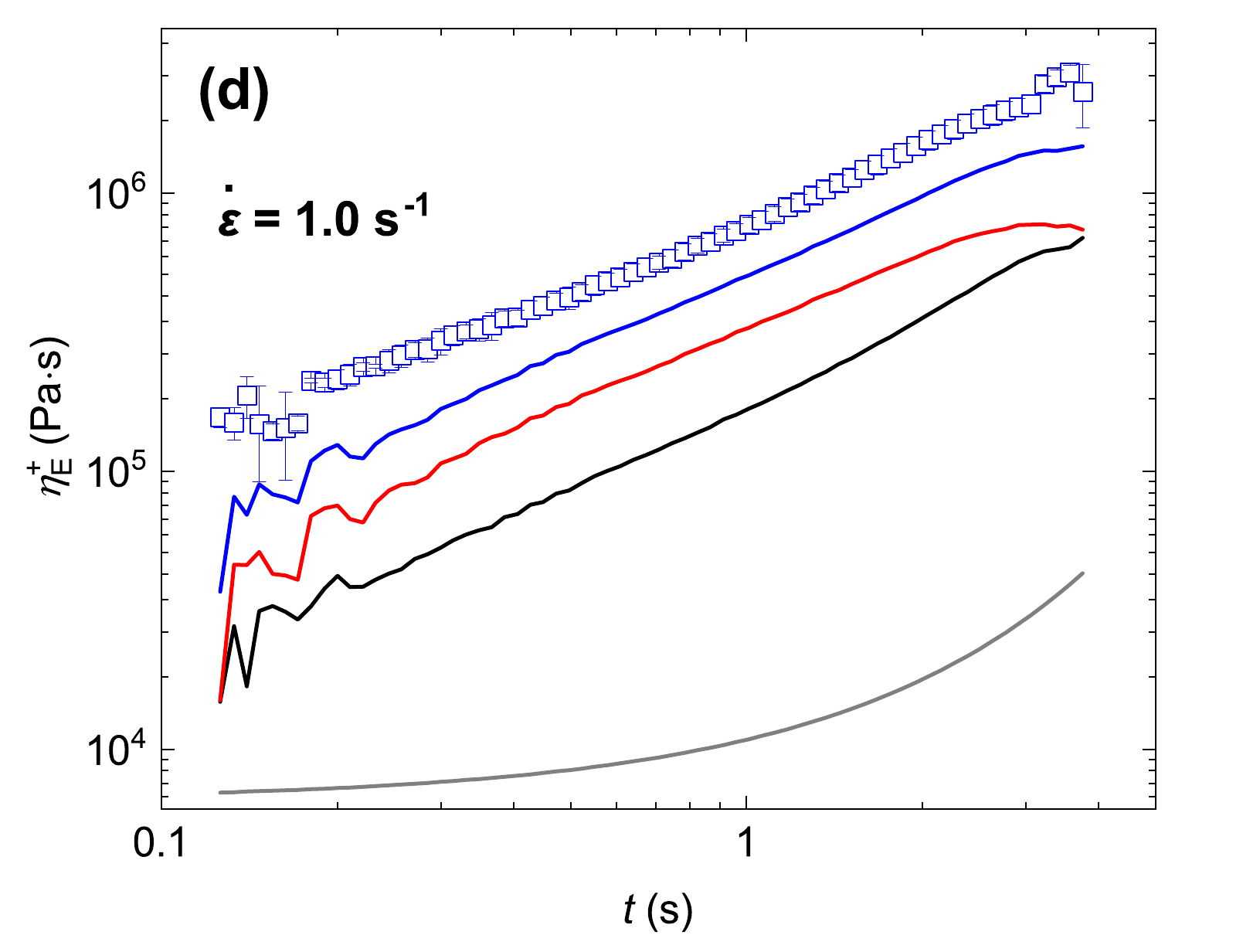}
  \includegraphics[width=8.6cm]{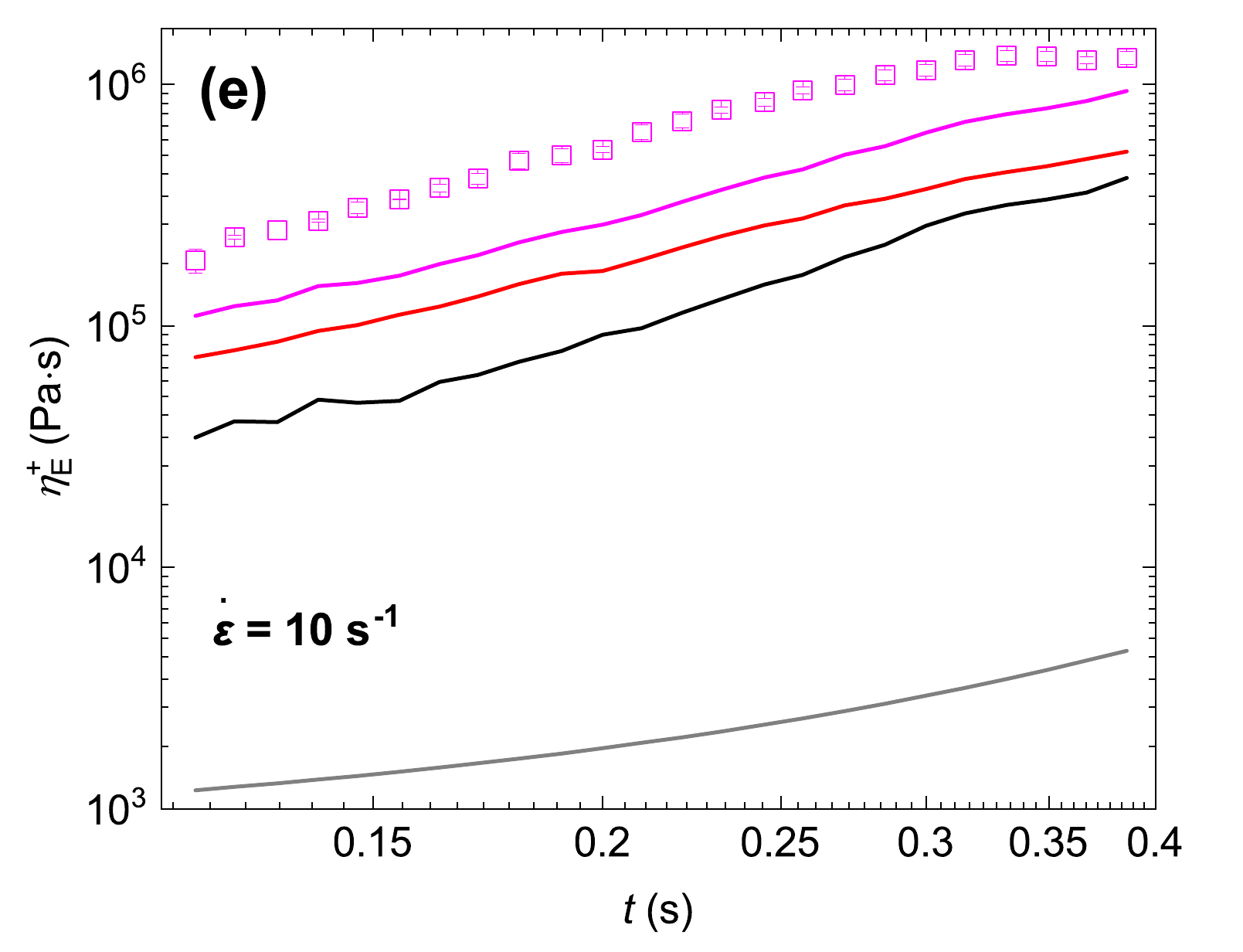}
  \includegraphics[width=8.6cm]{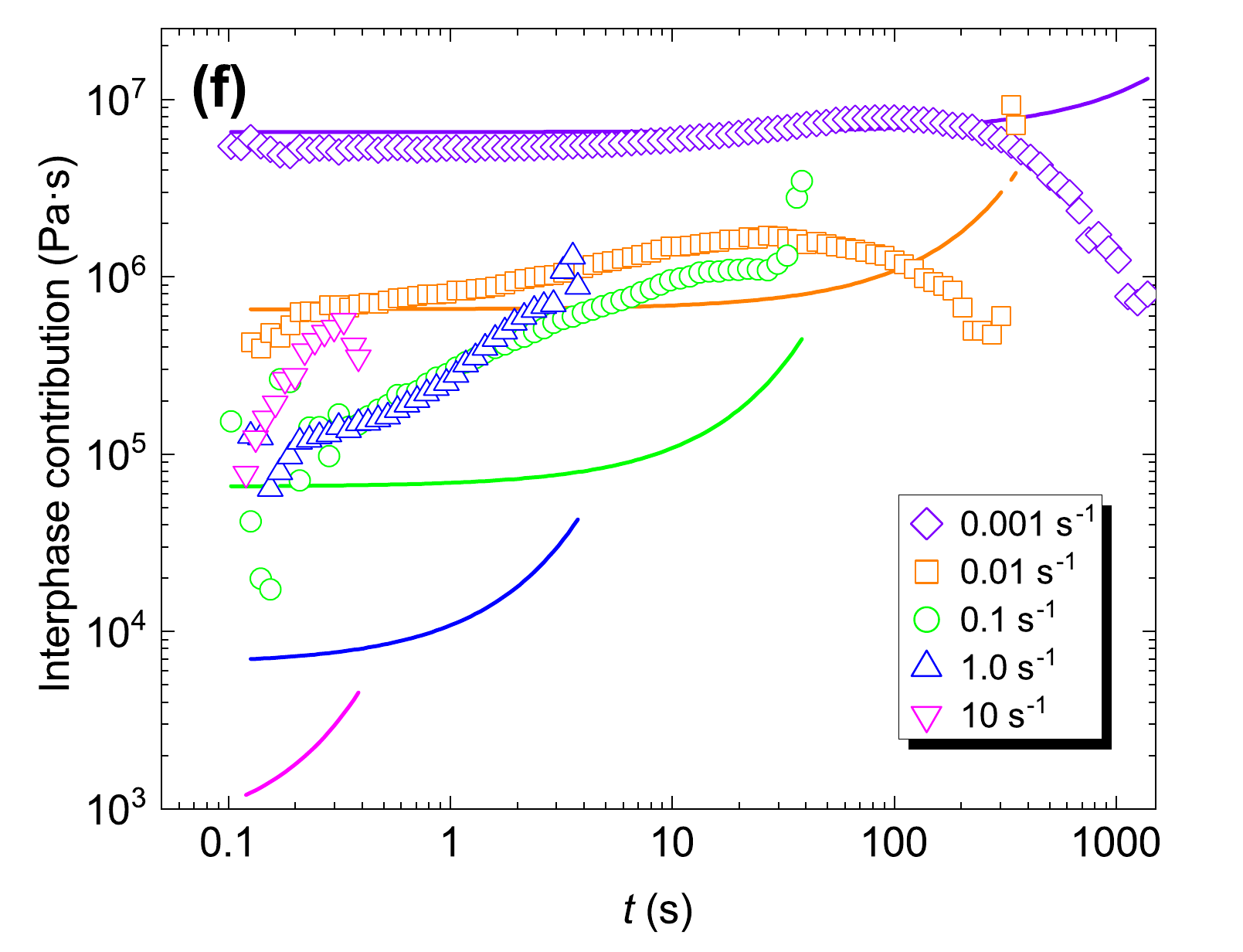}
  \caption{Comparison between experimental data (colored open symbols) and additivity rule with interphase for the film with 60/40 PS/PMMA composition and 2049 layers. The solid black, red, and grey lines represent each contribution from the model (PS, PMMA and interphase, respectively), while the colored line is the sum of these three contributions. Each Figure presents experiments done at a different strain rate: (a) 0.001, (b) 0.01, (c) 0.1, (d) 1 and (e) 10 $s^{-1}$. (f) presents the comparison between the theoretical (solid lines) and measured values of interphase contribution (colored open symbols).}
  \label{fig:fig4}
\end{figure*}

The model predictions with no adjustable parameter are then compared to the experimental measurements for a chosen film (60/40 PS/PMMA composition, 2049 layers) in Figure \ref{fig:fig4} (see Figure S7 for the other composition). First, we should note that this approach is different from the one proposed by Jordan \cite{Jordan2019}, in which $\Gamma$ was a fitting parameter. Second, we can verify that for films with small number of layers, the interfacial contribution is negligible and only improves marginally the fitting of the data (see Figure S8).
It is seen in Figure \ref{fig:fig4} that at low strain rates, 0.001 s$^{-1}$ and 0.01 s$^{-1}$, the experimental values are in a good agreement with the model except close to the breaking point (Figure \ref{fig:fig4}a, b) since the model does not predict strain-softening but, on the contrary, a slight strain-hardening due to an increase of the interfacial contribution related to the thinning of the sample over time.  With increasing strain rate, an increasing deviation from the experimental value is observed. The model predicts that the interphase contribution varies inversely proportional to the strain rate which leads to significant underestimates of the extensional viscosity of the multilayer films at strain rates above 0.1 s$^{-1}$.

We can then try to use the additivity rule with interphase to estimate what would be the necessary contribution to match the experimental data, by subtracting the response of PS and PMMA to the multilayer one.
Figure \ref{fig:fig4}f presents the comparison of the interphase contribution calculated from eq \ref{eq8} and extracted from our experimental results. As anticipated from the previous discussion, the contributions are similar at low strain rates and deviate strongly from each other at high strain rates. While under the hypotheses of the model the interphase contribution is inversely proportional to strain rate, the experimental contribution decreases less sharply and reaches similar viscosity values on the order of $10^5$ to $10^6$ Pa$\cdot$s for strain rates higher than 0.1 s$^{-1}$. Similar conclusions can be drawn from the results presented in Figure S7 concerning the 30/70 PS/PMMA composition. 

\subsubsection*{Interphase properties}

\begin{figure}[htb!]
  \centering
  \includegraphics[width=8.4cm]{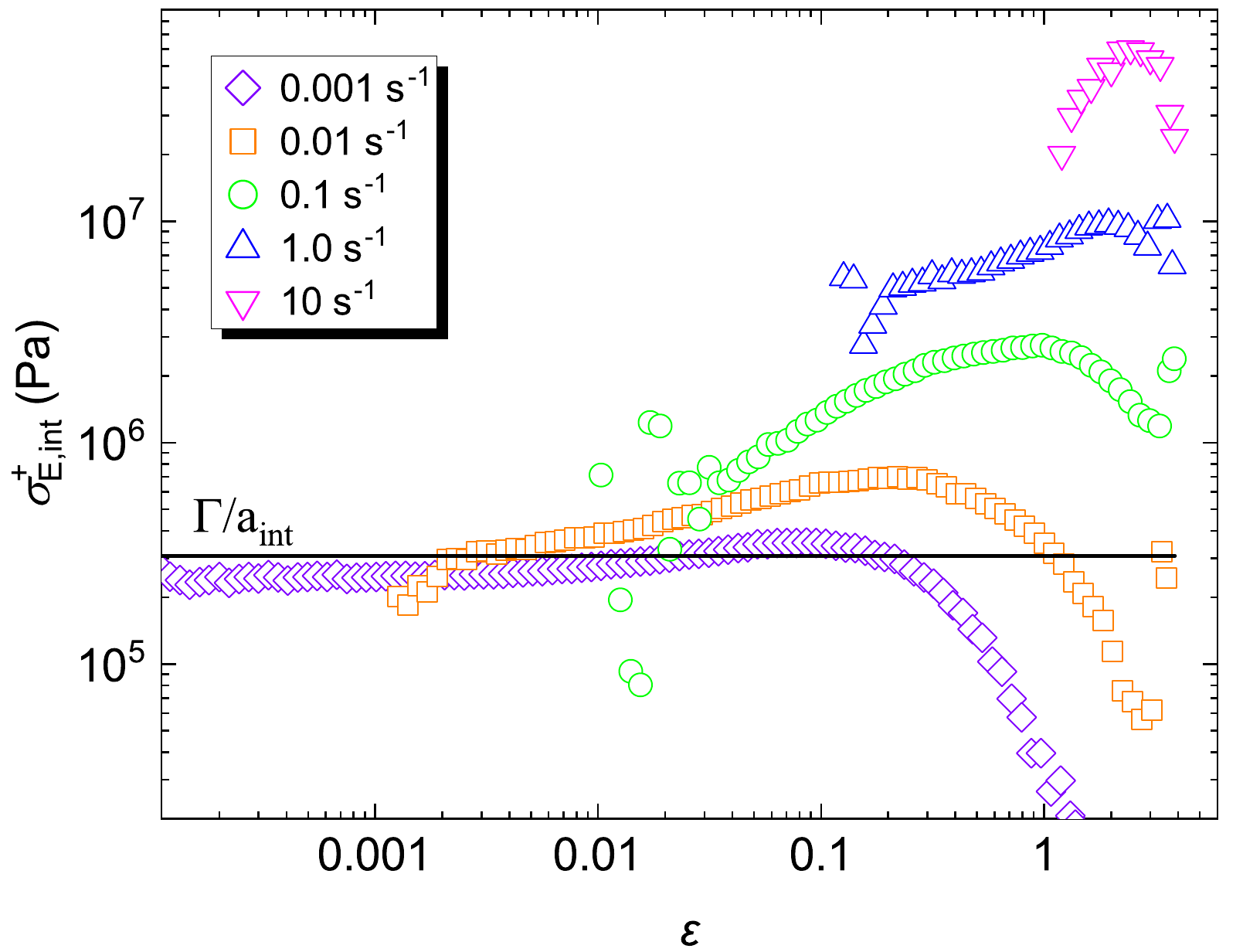}
  \caption{Measured interfacial stress as a function of strain for the same sample as in Figure \ref{fig:fig4}.} 
  \label{fig:fig5}
\end{figure}

\begin{figure}[hbt!]
  \centering
  \includegraphics[width=8.4cm]{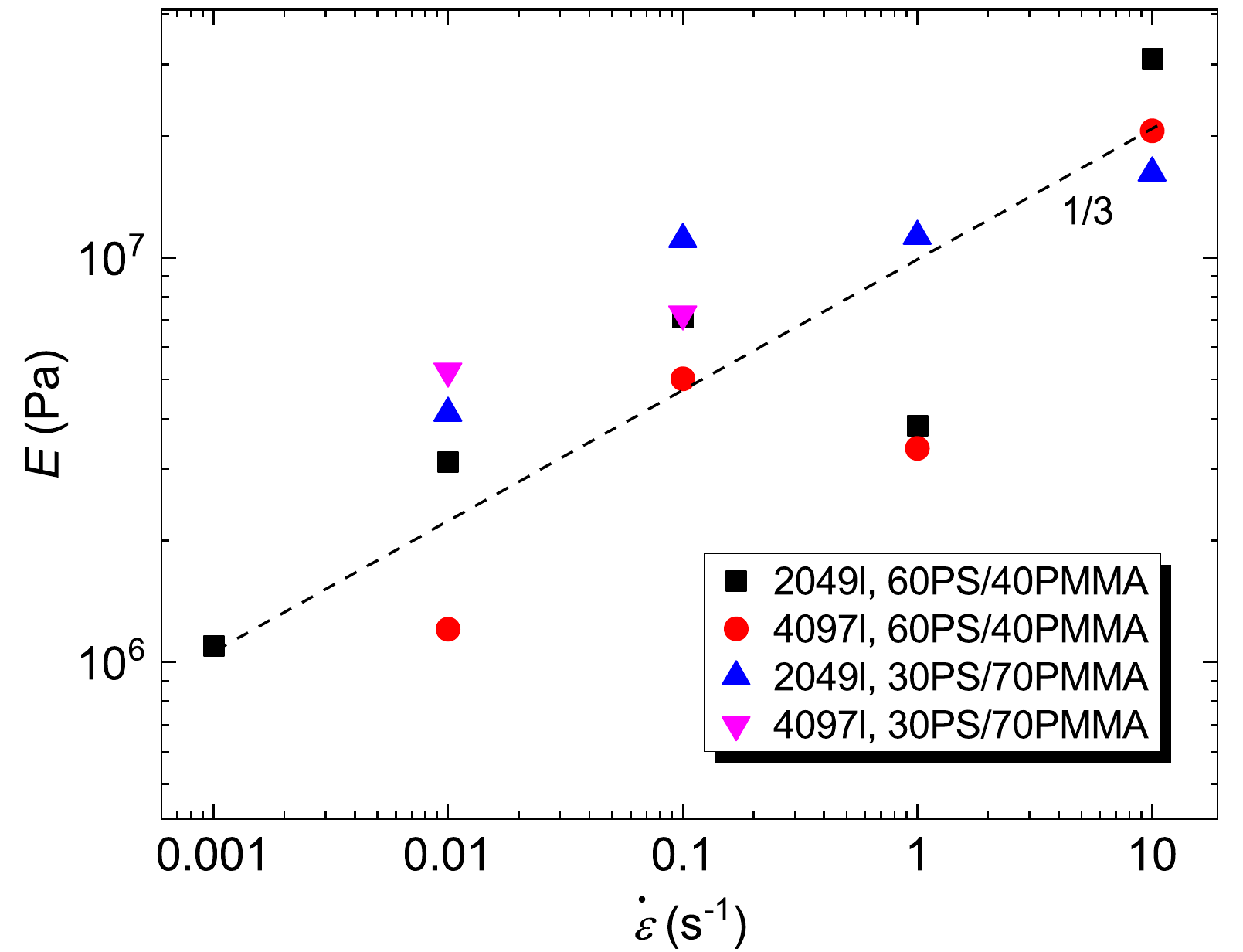}
  \caption{Interphase modulus as a function of strain rate for 2049 and 4097 layers at 60/40 PS/PMMA and 30/70 PS/PMMA compositions.} 
  \label{fig:fig6}
\end{figure}

To go further, let us study the interfacial stress extracted from the experimental data presented in Figure \ref{fig:fig4} and by using eqs \ref{eq6} and \ref{eq9}. The results for the same film as previously are presented as a function of strain, for each strain rate, in Figure \ref{fig:fig5} (interfacial stress for the film with 4097 layers and the other composition is presented on Figure S9).
At thermodynamic equilibrium, the interfacial stress shall be constant, no matter what are the strain and strain rate applied, and follow eq \ref{eq8}. Note that combining eqs \ref{eq8}, \ref{eq5} and \ref{eq10}, this interfacial stress could be expressed as a function of the Flory interaction parameter and Kuhn length only:
\begin{equation}
\sigma_{\text{\tiny{E,int,0}}}^+ \!\left(t\right) = \frac{kT\chi}{2b^3} \approx 300~ \text{kPa}
\label{eq11}
\end{equation}
which can be understood as an energy density of the monomers in the interphase.  
From Figure \ref{fig:fig5}, it is seen that this thermodynamic description of the interfacial stress describes well the experimental data at low strains and strain rates. However, there is an increase in the interfacial stress as strain increases above a critical value $\varepsilon_{\text{c}}$ close to 0.01, which becomes more pronounced at higher strain rates. This increase in stress may be due to the fact that the extensional flow modifies the interphase from its equilibrium conformation, as the chains become oriented at high strains or at strain rates such as $W_\text{i}>1$. In bulk, the elasticity of polymer melts and solutions in shear flows manifests itself through the existence of two non-zero normal stress differences. It is thus tempting to say that accordingly, when subjected to strong elongations, the interphase response becomes non-isotropic: the tensile stress is 2-dimensional, analogous of this difference in normal stress, \textit{i.e.} an anisotropy of the surface tension in the film. This anisotropy of the surface tension with respect to the direction is the signature of the 2D elasticity of the interphase \cite{miller_interfacial_2019}. A similar phenomenon occurs in soap films, where the Gibbs-Marangoni surface elasticity is due to the dilution of the surfactant at the interface and responsible for their stability \cite{gibbs_scientific_1961}.\\ 
If we consider the region above the critical strain, there is first a linear increase of the interphase contribution with strain, followed by a plateau and then a decrease before failure. Let us focus on the linear increase region (see Figure S10), in which we can write the interfacial stress as:
\begin{equation}
\sigma_{\text{\tiny{E,int}}}^+ \!\left(\varepsilon\right)=\sigma_{\text{\tiny{E,int,0}}}^+  + E_{\dot{\varepsilon}}\left(\varepsilon-\varepsilon_{\text{c}}\right)
\label{eq12}
\end{equation}
where the slope of the linear interfacial stress-strain region can then be termed an `interphase modulus', $E_{\dot{\varepsilon}}$ by analogy with the Gibbs-Marangoni surface elasticity. Values from about 1 to $\sim$10 MPa, which are typical of a rubbery plateau modulus, are obtained for $E_{\dot{\varepsilon}}$ as the strain rate increases from 0.001 to 10 s$^{-1}$.
Since this anisotropy is a dynamical effect, we observe a modulus which depends on the strain rate. Plotting this dependence of the modulus in Figure \ref{fig:fig6}, we can see that similar values of modulus are obtained at each strain rate for all samples and compositions, with a power-law dependence with strain rate having an exponent of about 1/3.
As stated previously, this interphase modulus shall be related to conformational changes\cite{Cunha2020,Zhang2022} close to the interface appearing at large strains and strain rates, but the precise description of the molecular dependency with respect to the strain rate is out of the scope of this paper.

\section*{Conclusion}
A systematic study of extensional properties of multilayer films of PS/PMMA, an immiscible polymer pair, has been conducted to probe the interphase rheological properties. A simple additivity rule based on summation of forces within PS and PMMA captures well the response from multilayer films with up to 129 layers. As the number of layers increases, the volume fraction of interphase becomes non negligible. A refined model proposed by Jordan et al. \cite{Jordan2019}, which includes an interfacial contribution in the additivity rule, captures well the behavior of our systems with more than 2000 layers and at low strain rate. At high strain rates however, the model underestimates the contribution of the interphase.
Looking at the interfacial stress, a deviation from thermodynamic equilibrium value related to interfacial tension and interphase thickness is observed at strains above a critical value of about 1 \%. A linear increase with strain is observed, with a slope increasing with increasing strain rate, leading to the measurement of an `interphase modulus' with values ranging from about 1 to several 10s of MPa. These values are those of a typical rubbery plateau modulus despite the fact that such non-compatibilized interphase are unentangled, suggesting a different behavior from the bulk and reminiscent of interfacial phenomena such as Gibbs-Marangoni elasticity. With this study, it is shown that extensional viscosity measurements can be used as a probe of determining the intrinsic `2D' rheological properties of interphases (\textit{i.e.} interfacial rheology), even for non-compatibilized systems. Having evidenced a surface elasticity occurring at high strain rates, it could be relevant to study its impact on the stability of nanolayers in elongational flows during processing such as nanolayer coextrusion.     

\section*{Acknowledgements}
The authors deeply acknowledge Frédéric Restagno for helpful suggestions on data analysis and interpretation. We also thank him and Ilias Iliopoulos for critical reading of the manuscript. The Ecole Doctorale SMI (ED 432) is acknowledged for granting A. D. the fellowship for her PhD work.

\bibliography{references}

\end{document}